\documentclass[twocolumn]{article}

\usepackage[utf8]{inputenc}

\usepackage{microtype}

\usepackage[numbers,square]{natbib}

\usepackage{color}
\usepackage{url}
\usepackage{booktabs}
\usepackage{hyperref}
\usepackage{amsmath}
\usepackage{nicefrac}
\usepackage{graphicx}
\usepackage{newtxmath}
\usepackage{bm}
\usepackage{flushend}
\usepackage{fancyhdr}

\newcommand{\y}{\bm{y}}
\newcommand{\Y}{\mathcal{Y}}
\renewcommand{\L}{\mathcal{L}}

\title{On the Futility of Learning Complex Frame-Level\\Language Models for Chord Recognition}

\author{Filip Korzeniowski and Gerhard Widmer\\
        {\small Department of Computational Perception}\\
        {\small Johannes Kepler University, Linz, Austria}\\
{\small \texttt{filip.korzeniowski@jku.at}}}
\date{}

\begin{document}

\maketitle 
\thispagestyle{fancy}

\begin{abstract}
Chord recognition systems use temporal models to post-process frame-wise chord
predictions from acoustic models. Traditionally, first-order models such as
Hidden Markov Models were used for this task, with recent works suggesting to
apply Recurrent Neural Networks instead. Due to their ability to learn
longer-term dependencies, these models are supposed to learn and to apply
musical knowledge, instead of just smoothing the output of the acoustic model.
In this paper, we argue that learning complex temporal models at the level of 
audio frames is futile on principle, and that non-Markovian models do not perform
better than their first-order counterparts. We support our argument 
through three experiments on the McGill Billboard dataset. The first
two show 1) that when learning complex temporal models at the frame level, 
improvements in chord sequence modelling are marginal; and 2) that these
improvements do not translate when applied within a full chord recognition
system. The third, still rather preliminary experiment gives first
indications that the use of complex sequential models for chord prediction at
higher temporal levels might be more promising.
\end{abstract}

\section{Introduction}

Computational systems that extract high-level information from signals face
two key problems: 1) how to extract meaningful information from noisy sources,
and 2) how to process and combine this information into sensible output.
For example, in domains such as speech recognition, these translate to
\emph{acoustic modelling} (how to predict phonemes from audio) 
and \emph{language modelling} (how to connect these phonemes into words and
sentences).

A similar structure can be observed in many systems related to music
processing, such as chord recognition. Chord recognition systems aim at
recognising and transcribing musical chords from audio, highly
descriptive harmonic features of music that form the basis of a myriad
applications (e.g.\ key estimation, cover song detection, or directly to
automatically provide lead sheets for musicians, to name a few). We refer
the reader to \cite{mcvicar_automatic_2014,cho_relative_2014} for an overview.

Chord recognition systems comprise two main parts: an \emph{acoustic model}
that extracts frame-wise harmonic features and predicts chord label
distributions for each time frame, and a \emph{temporal model} that connects
consecutive predictions and outputs a sequence of chord labels for a piece of
audio.

The temporal model provides coherence to the possibly volatile predictions of
the acoustic model. It also permits the introduction of higher-level musical
knowledge---we know from music theory that certain chord progressions are more
likely than others---to further improve the obtained chord labels.

A number of works (e.g.\ \cite{ni_endtoend_2012,pauwels_combining_2014,mauch_simultaneous_2010}) implemented hand-designed temporal models for chord
recognition. These models are usually first-order Dynamic Bayesian Networks
that operate at the beat or time-frame level. They are designed to incorporate
musical knowledge, with parameters set by hand or trained from data. 

A different approach is to learn temporal models fully from data, without any
imposed structure. Here, it is common to use simple Hidden Markov Models
\cite{cho2010exploring} or Conditional Random Fields
\cite{korzeniowski_fully_2016} with states corresponding to chord labels.

However, recent research showed that first-order models have very limited
capacity to to encode musical knowledge and focus on ensuring stability between
consecutive predictions (i.e.\, they only smooth the output sequence)
\cite{cho_relative_2014,chen_chord_2012}. In \cite{cho_relative_2014},
self-transitions dominate other transitions by several orders of magnitude, and
chord recognition results improve as self-transitions are amplified manually.
In \cite{chen_chord_2012}, employing a first-order chord transition model
hardly improves recognition accuracy, given a duration model is applied.

This result bears little surprise: firstly, many common chord patterns in pop,
jazz, or classical music span more than two chords, and thus cannot be
adequately modelled by first-order models; secondly, models that operate
at the frame level by definition only predict the chord symbol of the next
frame (typically ~10ms away), which most of the time will be the same as the
current chord symbol.

To overcome this, a number of recent papers suggested to use Recurrent Neural
Networks (RNNs) as temporal models\footnote{Note of our use of the term
``temporal model'' instead of ``language model'' as used in many
publications---the reason for this distinction will hopefully become clear at
the end of this paper.} for a number of music-related tasks, such as chord
recognition \cite{boulanger-lewandowski_audio_2013,sigtia_audio_2015} or
multi-f0 tracking \cite{sigtia_endtoend_2016}. RNNs are capable of modelling
relations in temporal sequences that go beyond simple first-order connections.
Their great modelling capacity is, however, limited by the difficulty to
optimise their parameters: exploding gradients make training instable, while
vanishing gradients hinder learning long-term dependencies from data.

In this paper, we argue that (neglecting the aforementioned problems) adopting
and training complex temporal models \emph{on a time-frame basis} is futile
\emph{on principle}. We support this claim by experimentally showing that they
do not outperform first-order models (for which we know they are unable to
capture musical knowledge) as part of a complete chord recognition system, and
perform only negligibly better at modelling chord label sequences. We thus
conclude that, despite their greater modelling capacity, the input
representation (musical symbols on a time-frame basis) prohibits learning and
applying knowledge about musical structure---the language models hence resort
to what their simpler first-order siblings are also capable of: smoothing the
predictions.

In the following, we describe three experiments: in the first, we judge two
temporal models directly by their capacity to model frame-level chord
sequences; in the second, we deploy the temporal models within a fully-fledged
chord recognition pipeline; finally, in the third, we learn a \emph{language
model} at chord level to show that RNNs are able to learn musical structure if
used at a higher hierarchical level. Based on the results, we then draw
conclusions for future research on temporal and language modelling in the
context of music processing.

\section{Experiment 1: Chord Sequence Modelling}

In this experiment, we want to quantify the modelling power of temporal models
directly. A temporal model predicts the next chord symbol in a sequence given
the ones already observed. Since we are dealing with frame-level data and adopt
a frame rate of 10 fps, a chord sequence consists of 10 chord symbols per
second. More formally, given a chord sequence $\y = (y_1,\ldots,y_{K})$, a
model $M$ outputs a probability distribution $P_M(y_k \mid y_1,\ldots,y_{k-1})$
for each $y_k$. From this, we can compute the probability of the chord sequence
\begin{align}
P_M\left(\y\right) = P_M(y_1) \cdot \Pi_{k=2}^{K} P_M\left(y_k \mid y_1, \ldots, y_{k - 1}\right).
\label{eq:seq_prod}
\end{align}
To measure how well a model $M$ predicts the chord sequences in a dataset, we
compute the average log-probability that it assigns to the sequences $\y$ of a
dataset $\Y$:
\begin{align}
        \L(M,\Y) = \frac{1}{N_\Y} \sum_{\y \in \Y} \log\left(P_M\left(\y\right)\right),
\end{align}
where $N_\Y$ is the total number of chords symbols in the dataset.

\subsection{Temporal Models}

We compare two temporal models in this experiment: A first-order Markov Chain
and a RNN with Long-Short Term Memory (LSTM) units. 

For the \textbf{Markov Chain}, $P_M(y_k \mid y_1, \ldots, y_{k-1})$ in
Eq.~\ref{eq:seq_prod} is simplified to $P_M(y_k \mid y_{k-1}) =
A_{y_k,y_{k-1}}$ due to the Markov property, and $P_M(y_1) =
\pi_{y_1}$. Both $\bm{\pi}$ and $\bm{A}$ can be estimated by
counting the corresponding occurrences in the training set.

For the \textbf{LSTM-RNN}, we follow closely the design, parametrisation and
training procedure proposed in \cite{sigtia_audio_2015}, and we refer the
reader to their paper for details. The input to the network at time step $k$ is
the chord symbol $y_{k-1}$ in one-hot encoding, the output is the probability
distribution $P_M(y_k \mid y_1, \ldots, y_{k-1})$ used in
Eq.~\ref{eq:seq_prod}. (For $k=1$, we input a ``no-chord'' symbol and the
network computes $P(y_1)$.) As loss we use the categorical cross-entropy
between the output distribution and the one-hot encoding of the target chord
symbol.

We use 2 layers of 100 LSTM units each, and add skip-connections such that
the input is connected to both hidden layers and to the output. We train the
network using stochastic gradient descent with momentum for a maximum of 200
epochs (the network usually converges earlier) with the learning rate
decreasing linearly from 0.001 to 0.  As in \cite{sigtia_audio_2015}, we show
the network sequences of 100 symbols (corresponding to 10 seconds) during
training. We experimented with longer sequences (up to 50 seconds), but results
did not improve (i.e.\, the network did not profit from longer contexts).
Finally, to improve generalisation, we augment the training data by randomly
shifting the key of the sequences each time they are shown during training.

\subsection{Data}

We evaluate the models on the McGill Billboard dataset
\cite{burgoyne_expert_2011}. We use songs with ids smaller than 1000 for
training, and the remaining for testing, which corresponds to the test protocol
suggested by the website accompanying the
dataset\footnote{\url{http://ddmal.music.mcgill.ca/research/billboard}}.  To
prevent train/test overlap, we filter out duplicate songs. This reduces the
number of pieces from 890 to 742, of which 571 are used for training and
validation (59155 unique chord annotations), and 171 for testing
(16809 annotations). 

The dataset contains 69 different chord types. These chord types are, to no
surprise, distributed unevenly: the four most common types (major, minor,
dominant~7, and minor~7) already comprise 85\% of all annotations. Following
\cite{cho_relative_2014}, we simplify this vocabulary to major/minor chords only,
where we map chords that have a minor 3rd as their first interval to minor, and
all other chords to major. After mapping, we have 24 chord symbols
($12\enspace\text{root notes} \times \{\text{major}, \text{minor}\}$) and a
special ``no-chord'' symbol, thus 25 classes.

\subsection{Results}

\begin{table}[]
\centering
\begin{tabular}{rcc}
\hline
                       & \textbf{Markov Chain} & \textbf{Recurrent NN} \\ \hline
$\L(M, \Y)$   & -0.273                 & -0.266                \\
$\L_c(M, \Y)$ & -5.420                  & -5.219                  \\
$\L_s(M, \Y)$ & -0.044                & -0.046                \\ \hline
\end{tabular}
\caption{Average log-probabilities of chords changes in the test set for the
        two temporal models. $\L(M, \Y)$ is the number for all
        chord symbols, $\L_c(M, \Y)$ for positions in the sequence
        where the chord symbol changes, and $\L_s(M, \Y)$ where
        it stays the same}
\label{tab:res_chordmod}
\end{table}

Table \ref{tab:res_chordmod} shows the resulting avg.\ log-probabilities of the
models on the test set. Additionally to $\L(M, \Y)$, we report
$\L_s(M, \Y)$ and $\L_c(M, \Y)$. These numbers represent the
average log-probability the model assigns to chord symbols in the dataset when
the current symbol is the \emph{same} as the previous one and, when it
\emph{changed}, respectively. They are computed similarly to $\L(M,
\Y)$, but the product in Eq.~\ref{eq:seq_prod} only captures $k$ where $y_k =
y_{k-1}$ or $y_k \neq y_{k-1}$, respectively.

They permit us to reason about how well a model will smooth the predictions
when the chord is stable, and how well it can predict chords when they change
(this is where ``musical knowledge'' could come into play).

We can see that the RNN performs only slightly better than the Markov Chain
(MC), despite its higher modelling capacity. This improvement is rooted in
better predictions when the chord changes (-5.22 for the RNN vs. -5.42 for the
MC).  This might indicate that the RNN is able to model musical knowledge
better than the MC after all. However, this advantage is minuscule and comes
seldom into play: the correct chord has an avg.\ probability of 0.0054 with the
RNN vs.  0.0044 with the MC\footnote{Both are worse than the random chance of
$\nicefrac{1}{25} = 0.04$, because both would still favour self-transitions},
and the number of positions at which the chord symbol changes, compared to
where it stays the same, is low.

In the next experiment, we evaluate if the marginal improvement provided
by the RNN translates into better chord recognition accuracy when deployed
in a fully-fledged system.

\section{Experiment 2: Frame-Level Chord Recognition}

In this experiment, we want to evaluate the temporal models in the context of a
complete chord recognition framework. The task is to predict for each audio
frame the correct chord symbol. We use the same data, the same train/test
split, and the same chord vocabulary (major/minor and ``no chord'') as
in Experiment 1.

Our chord-recognition pipeline comprises spectrogram computation, an
automatically learned feature extractor and chord predictor, and finally the
temporal model. The first two stages are based on our previous work
\cite{korzeniowski_fully_2016,korzeniowski_feature_2016}. We extract a
log-filtered and log-scaled spectrogram between 65 and 2100 Hz at 10 frames per
second, and feed spectral patches of 1.5s into one of three acoustic models: a
\emph{logistic regression classifier} (LogReg), a \emph{deep neural network}
(DNN) with 3 fully connected hidden layers of 256 rectifier units, and a
\emph{convolutional neural network} (ConvNet) with the exact architecture we
presented in \cite{korzeniowski_fully_2016}.

Each acoustic model yields frame-level chord predictions, which are then
processed by one of three different temporal models. 

\subsection{Temporal Models}

We test three temporal models of increasing complexity.  The simplest one is
Majority Voting (MV) within a context of 1.3s, The others are the very same we
used in the previous experiment.

Connecting the Markov Chain temporal model to the predictions of the acoustic
model results in a \emph{Hidden Markov Model} (HMM). The output chord sequence
is decoded using the Viterbi algorithm.

To connect the RNN temporal model to the predictions of the acoustic model,
we apply the \emph{hashed beam search} algorithm, as introduced in
\cite{sigtia_audio_2015}, with a beam width of 25, hash length of 3 symbols
and a maximum of 4 solutions per hash bin. The algorithm only approximately
decodes the chord sequence (no efficient and exact algorithms exist, because
the output of the network depends on \emph{all} previous inputs).

\subsection{Results}

\begin{table}[]
\centering
\begin{tabular}{@{}lllll@{}}
\toprule
                 & \textbf{None} & \textbf{MV} & \textbf{HMM} & \textbf{RNN} \\ \midrule
\textbf{LogReg}  & 72.3          & 72.8        & 73.4         & 73.1         \\
\textbf{DNN}     & 74.2          & 75.3        & 76.0         & 75.7         \\
\textbf{ConvNet} & 77.6          & 78.1        & 78.9         & 78.7         \\ \bottomrule
\end{tabular}
\caption{Weighted Chord Symbol Recall of the 24 major and minor chords and the
         ``no-chord'' class for the tested temporal models (columns) on
         different acoustic models (rows).}
\label{tab:res_chordrec}
\end{table}

Table \ref{tab:res_chordrec} shows the Weighted Chord Symbol Recall (WCSR) of
major and minor chords for all combinations of acoustic and temporal models.
The WCSR is defined as \(\mathcal{R} = \nicefrac{t_{c}}{t_{a}}\), where $t_c$
is the total time where the prediction corresponds to the annotation, and $t_a$
is the total duration of annotations of the respective chord classes (major and
minor chords and the ``no-chord'' class, in our case). We used the
implementation provided in the ``mir\_eval'' library \cite{raffel_mir_2014}.

The results show that the complex RNN temporal model does not outperform
the simpler first-order HMM. They improve compared to not using a temporal
model at all, and to a simple majority vote.

The results suggest that the RNN temporal model does not display its
(marginal) advantage in chord sequence modelling when deployed within the
complete chord recognition system. We assume the reasons to be 1) that the
improvement was small in the first place, and 2) that exact inference is not
computationally feasible for this model, and we have to resort to approximate
decoding using beam search.

\section{Experiment 3: Modelling Chord-level Label Sequences}

\begin{figure*}[!ht]
        \centering
        \includegraphics[width=\textwidth]{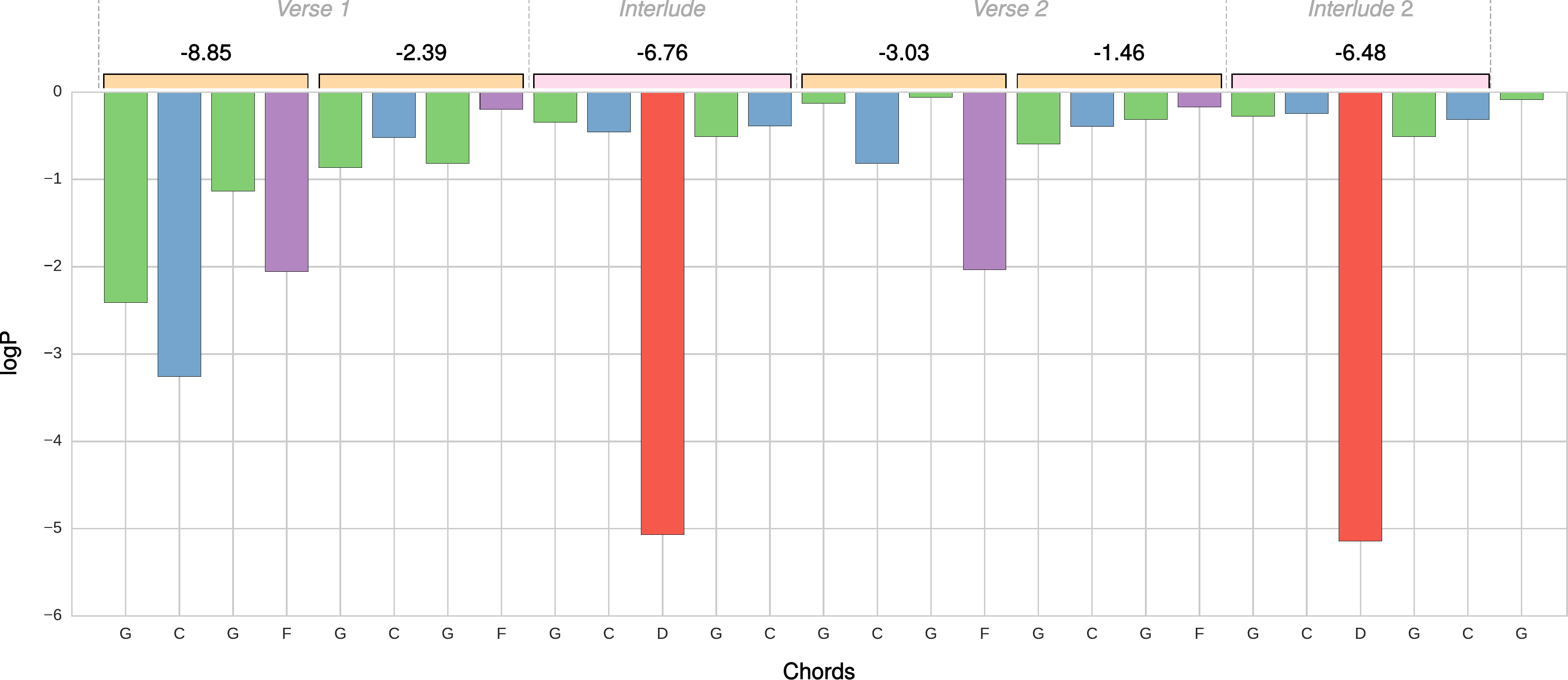}
        \caption{Log-probabilities of chords at the beginning of The Beatles'
                 ``A Hard Day's Night'', as computed by a RNN language model at
                 the chord level. Bar colors indicate chord type: green
                 corresponds to G major, blue to C major, purple to F major,
                 and red to D major.
                 We observe that the two repeated chord sequences (``G-C-G-F''
                 and ``G-C-D-G-C'', marked with light orange and light pink
                 on the top) achieve higher probabilities as they
                 are repeated, with the exception of one repetition of the
                 first sequence at the beginning of Verse 2 (in this case, the
                 network did not expect to see the ``F'' after several
                 repetitions of a G-C transition).  This indicates that the
                 network was able to remember to some degree the chord
                 progressions seen earlier in the song.}
        \label{fig:harddaysnight}
\end{figure*}

In the final experiment, we want to support our argument that the RNN does not
learn musical structure because of the hierarchical level (time frames) it is
applied on. To this end, we conduct an experiment similar to the first
one---an RNN is trained to predict the next chord symbol in the sequence.
However, this time the sequence is not sampled at frame level, but at chord
level (i.e.\ no matter how long a certain chord is played, it is reduced to a
single instance in the sequence). Otherwise, the data, train/test split, and
chord vocabulary are the same as in Experiment 1.

The results confirm that in such a scenario, the RNN clearly outperforms the
Markov Chain (Avg.\ Log-P.\ of -1.62 vs.\ -2.28). Additionally, we observe that
the RNN does not only learn static dependencies between consecutive chords; it
is also able to adapt to a song and recognise chord progressions seen
previously in this song without any on-line training. This resembles the way
humans would expect the chord progressions not to change much during a part
(e.g.\ the chorus) of a song, and come back later when a part is repeated.
Figure \ref{fig:harddaysnight} shows exemplary results from the test data.

\section{Conclusion}

We argued that learning complex temporal models for chord recognition
\footnote{We expect similar results for other music-related tasks.} on a
time-frame basis is futile. The experiments we carried out support our
argument. The first experiment focused on how well a complex temporal model
can learn to predict chord sequences compared to a simple first-order one. We
saw that the complex model, despite its substantially greater modelling
capacity, performed only marginally better. The second experiment showed that,
when deployed within a chord recognition system, the RNN temporal model did
not outperform the first-order HMM. Its slightly better capability to
model frame-level chord sequences was probably counteracted by the approximate
nature of the inference algorithm. Finally, in the third experiment, we
showed preliminary results that when deployed at a higher hierarchical level
than time frames, RNNs are indeed capable of learning musical structure
beyond first-order transitions.

Why are complex temporal models like RNNs unable to model frame-level chord
sequences? We believe the following two circumstances to be the causes:
1)~transitions are dominated by self-transitions, i.e.\ models need to predict
self-transitions as well as possible to achieve good predictive results on the
data, and 2)~predicting chord changes ``blindly'' (i.e.\ without knowledge
\emph{when} the change might occur) competes with 1) via the normalisation
constraint of probability distributions.

Inferring when a chord changes proves difficult if the model can only consider
the \emph{frame-level} chord sequence. There are simply too many uncertainties
(e.g.\ the tempo and harmonic rhythm of a song, timing deviations, etc.) that
are hard to estimate. However, the models also do not have access to features
computed from the input signal, which might help in judging whether a chord
change is imminent or not. Thus, the models are blind to the \emph{time} of a
chord change, which makes them focus on predicting self-transitions, as we
outlined in the previous paragraph.

We know from other domains such as natural language modelling that RNNs are
capable of learning state-of-the-art language models
\cite{mikolov_recurrent_2010}. We thus argue that the reason they underperform
in our setting is the frame-wise nature of the input data. For future research,
we propose to focus on \emph{language models} instead of frame-level temporal
models for chord recognition. By ``language model'' we mean a model at a higher
hierarchical level than the temporal models explored in this paper (hence the
distinction in name)---like the model used in the final experiment described in
this paper. Such language models can then be used in sequence classification
framework such as sequence transduction \cite{graves_sequence_2012} or
segmental recurrent neural networks \cite{lu_segmental_2016}.

Our results indicate the necessity of hierarchical models, as postulated in
\cite{widmer_getting_2016a}: powerful feature extractors may operate at the
frame-level, but more abstract concepts have to be estimated at higher temporal
(and, conceptual) levels. Similar results have been found for other
music-related tasks: e.g., in \cite{srinivasamurthy_generalized_2016},
dividing longer metrical cycles into their sub-components led to improved beat
tracking results; in the field of musical structure analysis (an obviously
hierarchical concept), \cite{mcfee_analyzing_2014} extracted representations
on different levels of granularity (although their system scored lower in
traditional evaluation measures for segmentation, it facilitates a hierarchical
breakdown of a piece).

Music theory teaches that cadences play an important role in the harmonic
structure of music, but many current state-of-the-art chord recognition systems
(including our own) ignore this. Learning powerful language models at sensible
hierarchical levels bears the potential to further improve the accuracy of
chord recognition systems, which has remained stagnant in recent years.

\section*{Acknowledgements}

This work is supported by the European Research Council (ERC) under the EU’s
Horizon 2020 Framework Programme (ERC Grant Agreement number 670035, project
``Con Espressione''). The Tesla K40 used for this research was donated by the
NVIDIA Corporation.

\bibliographystyle{jaes}

\bibliography{aes2017}

\end{document}